\documentclass[aip, jcp, notitlepage, floatfix, reprint, citeautoscript, twocolumn]{revtex4-1}

\usepackage{amsmath}
\usepackage{amsfonts}
\usepackage{graphicx}
\usepackage{dcolumn}
\usepackage[table]{xcolor}
\usepackage[version=3]{mhchem}
\usepackage{transparent}
\usepackage{gensymb}
\usepackage{tabularx}
\usepackage{etoolbox}
\usepackage[hidelinks]{hyperref}
\usepackage{bm}
\usepackage{relsize}
\usepackage[normalem]{ulem}
\usepackage{miller}

\def\maxfloatwidth{%
  \ifdim\columnwidth>246.0pt
  300.0pt  \else
  \columnwidth
  \fi
}

\newcommand{\mbf}[1]{\mathbf{#1}}
\newcommand{\tcr}[1]{\textcolor{black}{#1}}
\newcommand{\tcb}[1]{\textcolor{black}{#1}}
\newcommand{\etal}{\emph{et al.}}

\begin{document}

\title{Stabilization of AgI's polar surfaces by the aqueous
  environment, and its implications for ice formation}

\author{Thomas Sayer}
\affiliation{Department of Chemistry, University of Cambridge,
  Lensfield Road, Cambridge CB2 1EW, United Kingdom}

\author{Stephen J. Cox}
\email{sjc236@cam.ac.uk}
\affiliation{Department of Chemistry, University of Cambridge,
  Lensfield Road, Cambridge CB2 1EW, United Kingdom}

\date{\today}

\begin{abstract}
  Silver iodide is one of the most potent inorganic ice nucleating
  particles known, a feature generally attributed to the excellent
  lattice match between its basal \ce{Ag}-\hkl(0001) and
  \ce{I}-\hkl(000-1) surfaces, and ice. This crystal termination,
  however, is a type-III polar surface, and its surface energy
  therefore diverges with crystal size unless a polarity compensation
  mechanism prevails. In this simulation study, we investigate to what
  extent the surrounding aqueous environment is able to provide such
  polarity compensation. On its own, we find that pure \ce{H2O} is
  unable to stabilize the \ce{AgI} crystal in a physically reasonable
  manner, and that mobile charge carriers such as dissolved ions, are
  essential. In other words, proximate dissolved ions must be
  considered an integral part of the heterogeneous ice formation
  mechanism. The simulations we perform utilize recent advances in
  simulation methodology in which appropriate electric and electric
  displacement fields are imposed. A useful by-product of this study
  is the direct comparison to the commonly used Yeh-Berkowitz method
  that this enables. Here we find that naive application of the latter
  leads to physically unreasonable results, and greatly influences the
  structure of \ce{H2O} in the contact layer. We therefore expect
  these results to be of general importance to those studying
  polar/charged surfaces in aqueous environments.
\end{abstract}

\maketitle

\section{Introduction}
\label{sec:intro}

The formation of ice is one of the most prevalent and important phase
transitions on Earth. In sufficiently pure samples, water can exist in
a supercooled liquid state to temperatures as low as
approx. $-38^\circ$C \cite{rosenfeld2000deep,murray2010kinetics}. The
fact that ice formation is routinely observed close to the melting
temperature is due to a process known as \emph{heterogeneous
  nucleation}, whereby the presence of foreign bodies facilitates
crystallization. These foreign bodies are often referred to as
\emph{ice nucleating particles} (INPs) \cite{vali2015technical}, and
examples of particularly effective INPs include the bacterium
\emph{Pseudomonas syringae} \cite{maki1974ice,pandey2016ice},
cholesterol \cite{head1961steroids,sosso2018unravelling}, and feldspar
\cite{atkinson2013importance,harrison2016not,kiselev2017active}. Owing
to the importance of heterogeneous ice nucleation across a range of
fields from atmospheric chemistry \cite{murray2012ice} to cryobiology
\cite{morris2013controlled}, understanding the molecular mechanisms by
which such INPs promote ice formation is the frequent study of both
experiments
\cite{maki1974ice,head1961steroids,atkinson2013importance,harrison2016not,kiselev2017active,knopf2006heterogeneous,dymarska2006deposition,wilson2015marine,whale2015ice,niedermeier2010heterogeneous,hiranuma2015ice,marcolli2016ice,reischel1975freezing,evans1965requirements,anderson1976supersaturation,whale2018enhancement,kumar2018ice}
and simulations
\cite{sosso2016crystal,lupi2014heterogeneous,lupi2014does,lupi2016pre,zielke2015simulations,cox2013microscopic,sosso2016ice,cox2015molecularI,cox2015molecularII,fitzner2015many,hudait2018ice,glatz2017heterogeneous,bi2016heterogeneous,cabriolu2015ice,bi2017enhanced,sosso2018unravelling,zielke2016simulations,zielke2014molecular,fraux2014note,glatz2016surface}. The
INP we investigate here is \ce{AgI}, which is perhaps the most potent
inorganic INP currently known
\cite{marcolli2016ice,evans1965requirements,reischel1975freezing,vonnegut1947nucleation}. In
particular, we consider the basal Ag-\hkl(0001) and I-\hkl(000-1)
crystal faces---the focus of numerous
\cite{zielke2016simulations,zielke2014molecular,fraux2014note,glatz2016surface,hale1980studies,ward1983study,ward1982monte}
studies---and exploit recent advances in simulation methodology
\cite{stengel2009electric,zhang2016computing1,zhang2016finite,sprik2018finite}
to better understand plausible mechanisms by which the aqueous
environment can stabilize these interfaces.

The suggested reason for \ce{AgI}'s excellent ice nucleating ability
is often stated to be its close structural similarity to ice
\cite{pk97}. Indeed, it was this fact that first led Vonnegut
\cite{vonnegut1947nucleation} to test the efficacy of \ce{AgI} as an
INP. This rather appealing and intuitive suggestion of course
presupposes that the crystal structure, especially close to the
surface of the crystal, is stable in an aqueous environment. This is
not a trivial matter. The complicating factor arises from the wurtzite
structure of the \ce{AgI} crystal: When cleaved so as to expose its
Ag-\hkl(0001) and I-\hkl(000-1) faces, it forms a polar surface. (It
is a type-III polar surface in Tasker's classification
\cite{tasker1979stability}.)  If we assume that the \ce{Ag+} and
\ce{I-} ions occupy positions that closely resemble that of bulk
\ce{AgI}, a so-called `bulk termination', then arguments based on
classical electrostatics show that the electrostatic contribution to
the surface energy of the crystal diverges with the
width \footnote{\tcb{Throughout this article, we use `width' to refer
    to the crystal's extent along \hkl[0001].}} of the crystal
\cite{tasker1979stability,nosker1970polar}. Simply put, for crystals
thicker than a few atomic layers, this polar surface termination is
unstable. Thus, if the Ag-\hkl(0001) and I-\hkl(000-1) surfaces are to
promote ice formation by acting as a template, a stabilization
mechanism is required.

Polar surfaces similar to the Ag-\hkl(0001) and I-\hkl(000-1) surfaces
of \ce{AgI} are common in semiconductors and metal
oxides. Accordingly, there is a wide body of experimental and
theoretical work aimed at understanding the stabilization mechanisms
of such surfaces, which has been reviewed extensively by Noguera and
co-workers \cite{noguera2000polar,goniakowski2008polarity}. The
essential feature of any stabilization mechanism is \emph{polarity
  compensation}, where the presence of a compensating net charge (CNC)
at the interface ensures electrostatic stability. Further details
regarding polarity compensation are given in
\tcr{Sec.~\ref{sec:GenTheor}}. As discussed in
Refs.~\onlinecite{noguera2000polar,goniakowski2008polarity}, three
plausible mechanisms are: (i) electronic reconstruction i.e., partial
filling of electronic surface states; (ii) nonstoichiometric
reconstruction i.e., modification of the surface region's composition;
and (iii) adsorption of charged foreign species. This last mechanism
is of particular interest with regard to ice formation, as the aqueous
environment may be able to supply the required CNC, either from
dissolved ions, or from the dielectric properties of water
itself. Understanding polarity compensation from the aqueous
environment is therefore one of the central themes of this study.

Owing to its excellent ice nucleating properties, the
\ce{AgI}/\ce{H2O} interface has been the focus of many previous
studies
\cite{marcolli2016ice,reischel1975freezing,evans1965requirements,anderson1976supersaturation,zielke2016simulations,zielke2014molecular,fraux2014note,glatz2016surface,hale1980studies,ward1983study,ward1982monte}. From
a simulation perspective, however, it is only recently that
computational resources have been such that ice formation at \ce{AgI}
has been tackled directly. Zielke \emph{et al}
\cite{zielke2014molecular}, and Fraux and Doye investigated ice
formation at different crystal faces of \ce{AgI}
\cite{fraux2014note}. For the wurtzite structure considered here, both
sets of authors found that ice formation occurred at Ag-\hkl(0001),
and that no ice formation was observed at either the I-\hkl(000-1) or
\hkl(10-10) faces. This was attributed to the fact that the water in
contact with Ag-\hkl(0001) formed hexagonal rings that had a bilayer
structure similar to ice. On the other hand, although hexagonal rings
also formed at I-\hkl(000-1), these had a more coplanar structure, and
were less able to promote ice-like structures in the water more
distant from the interface. At the \hkl(10-10) interface, both studies
found no ice-like structures in the contact layer. Glatz and Sarupria
\cite{glatz2016surface} subsequently studied ice formation at
Ag-\hkl(0001), and found that changes in the charge distribution
within the crystal framework had significant effects on ice
formation. Consistent with Zielke \emph{et al}, and Fraux and Doye,
they found that \ce{AgI} facilitated ice formation by promoting
hexagonal ice-like structures in the contact layer.

While the work in
Refs.~\onlinecite{zielke2014molecular,fraux2014note,glatz2016surface}
have provided potential molecular mechanisms by which ice forms at
\ce{AgI}, they have assumed bulk termination of the crystal structure,
either by employing completely immobile \ce{AgI}, or by imposing
restraining potentials to the \ce{Ag+} and \ce{I-} ions so as to
maintain a structure close to that of the bulk crystal. Although Fraux
and Doye did attempt to use a classical force field to model the
motion of the \ce{AgI} crystal, they reported that the crystal quickly
dissolved. They also found that in order to observe ice formation,
unrealistically strong restraining potentials had to be imposed. When
the strength of the restraining potentials was reduced so that the
widths of the peaks in the bulk radial distribution function were
reproduced, no ice formation was observed. This state of affairs is
clearly far from ideal, and establishing simulation protocols to
tackle ice formation not only at polar surfaces, but also charged
interfaces in general, presents a significant advancement of the
field. This is especially timely given recent experimental studies
regarding the role of ions on heterogeneous ice nucleation
\cite{whale2018enhancement,kumar2018ice}.

In this study, the central issue that we seek to address is whether or
not an aqueous environment can provide adequate charge compensation
such that the structures of the Ag-\hkl(0001) and I-\hkl(000-1) faces
closely resemble their bulk terminations, and if so, what effect the
stabilization mechanisms have on ice formation at these interfaces. To
achieve this goal, we will exploit the finite field methods recently
developed in
Refs.~\onlinecite{zhang2016computing1,zhang2016finite,sprik2018finite}. We
will show that while the dielectric properties of water are in
principle sufficient to stabilize the \ce{AgI} crystal, this leads to
unphysically large fields in the fluid, which would likely result in
the dielectric breakdown of water. This problem is circumvented upon
the introduction of free ions in solution, which are able to stabilize
the crystal while maintaining zero average electric field in the
solution. When ice forms in this system, a proton ordered contact
layer is found at Ag-\hkl(0001). Whereas in the absence of free ions
this proton ordering persists far from the surface, coordination of
the water molecules to the ions is sufficient to destroy this proton
ordering beyond the contact layer.

The article is outlined as follows. First, we feel it is instructive
to give an account of the technical challenges faced when trying to
simulate polar systems such as \ce{AgI} in contact with water. In
Sec.~\ref{sec:GenTheor} we therefore present a comparison study of the
commonly used Yeh-Berkowitz \cite{YehBerkowitz1999sjc} method and the
finite field methods. This also provides a useful context in which to
provide the required background theory. In Sec.~\ref{subsec:IceNoIons}
we then go on to investigate ice formation in a system that comprises
pure water in contact with a slab of \ce{AgI} that is held fixed. The
purpose here is to compare the effects of different electrostatic
boundary conditions, which also allows us to compare to previous
studies
\cite{zielke2014molecular,fraux2014note,glatz2016surface}. Where
appropriate, we then extend these results to systems in which the
\ce{AgI} is allowed to move. We forewarn the reader that the results
presented in Sec.~\ref{subsec:IceNoIons} unlikely reflect an
experimentally realizable scenario; they are included for illustrative
and comparison purposes. In Sec.~\ref{subsec:IceWithIons} we present
the main results of this article, namely, the influence of dissolved
ions on the ice formation mechanism at \ce{AgI}. We summarize and
discuss future directions in Sec.~\ref{sec:concl}. Methods are
outlined in Sec.~\ref{sec:methods}.

\section{Using finite fields to model silver iodide crystals relevant to ice formation}
\label{sec:GenTheor}

\begin{figure}[tb]
  \includegraphics[width=7.65cm]{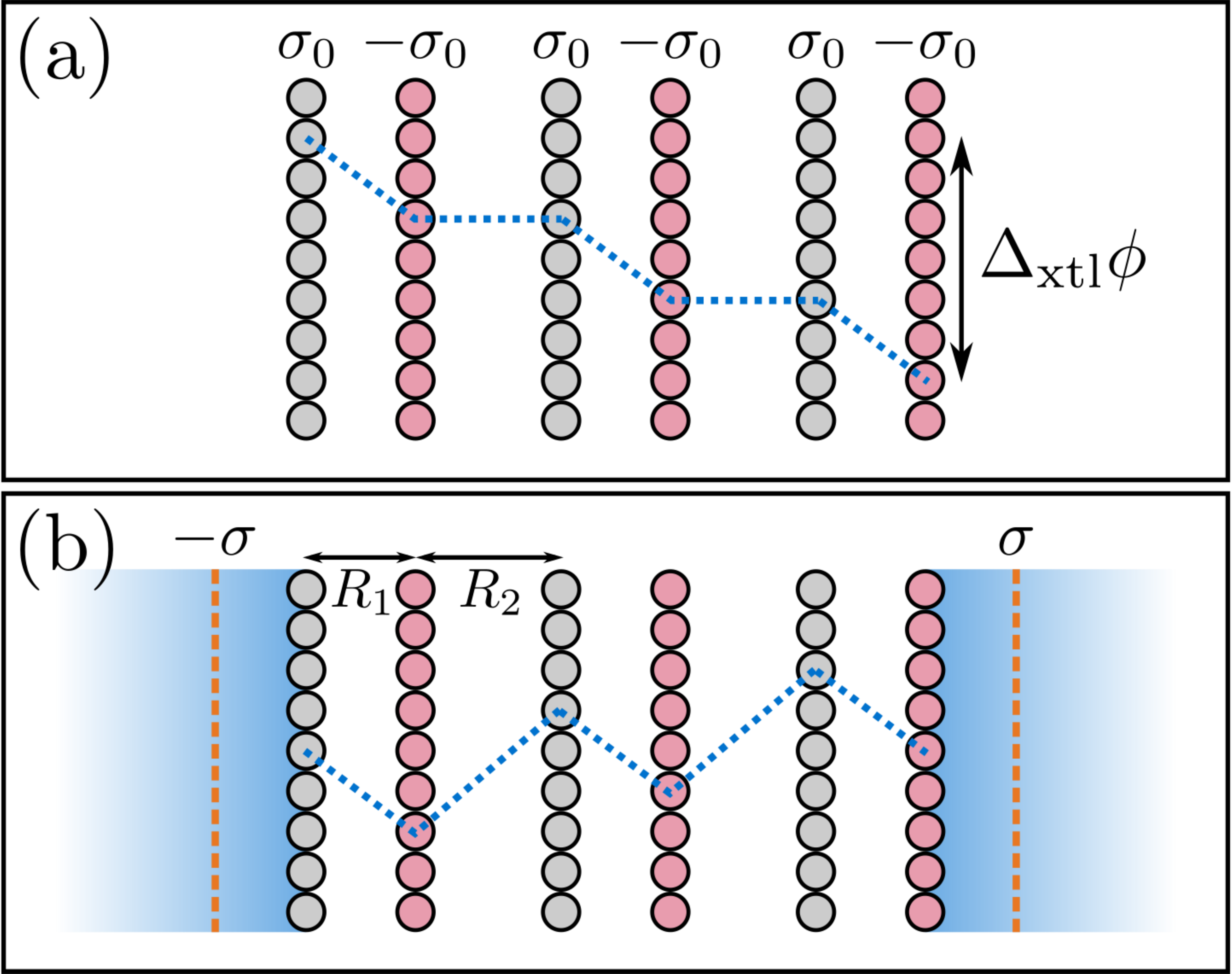}
  \caption{The unreconstructed Ag-\hkl(0001) and I-\hkl(000-1)
    surfaces require polarity compensation for large crystal
    sizes. (a) In vacuum, alternating layers of \ce{Ag+} (silver
    circles) and \ce{I-} ions (pink circles), respectively carrying
    surface charge densities $\sigma_0$ and $-\sigma_0$, establish a
    potential drop $\Delta_{\rm xtl}\phi$ across the
    slab. $\Delta_{\rm xtl}\phi$ increases linearly with the width of
    the slab, resulting in a divergence of the surface energy. (b) In
    an aqueous electrolyte solution, a Helmholtz layer (orange dashes)
    with surface charge density $\sigma$ is established, which reduces
    $|\Delta_{\rm xtl}\phi|$. If $\sigma = \sigma_{\rm CNC}$, then
    $\Delta_{\rm xtl}\phi = 0$, as shown. Blue dotted lines show
    schematic electrostatic potential profiles $\phi(z)$. The
    separations between crystal planes are given by $R_1$ and $R_2$.}
  \label{fig:PolarSlabSchematic}
\end{figure}

Particles of \ce{AgI} that promote ice formation typically have
diameters on the order $10^2-10^3$\,nm \cite{marcolli2016ice}. Along
any particular crystallographic direction, we may therefore expect to
encounter on the order of $10^3-10^4$ atomic layers. Such sizes are
sufficiently large that any \ce{AgI} crystals exposing their
Ag-\hkl(0001) and I-\hkl(000-1) faces must undergo some kind of
polarity compensation mechanism. This can be understood with the aid
of Fig.~\ref{fig:PolarSlabSchematic}, which shows a schematic of an
unreconstructed \ce{AgI} slab exposing its basal faces. Along this
crystallographic direction, the crystal comprises alternating layers
of \ce{Ag+} and \ce{I-} ions, each bearing a surface charge density of
$\sigma_0$ and $-\sigma_0$, respectively. In
Fig.~\ref{fig:PolarSlabSchematic}\,(a), the slab is surrounded on
either side by vacuum, and upon it we have superimposed a
representation of the electrostatic potential profile $\phi(z)$. It is
straightforward to infer from this that the potential drop across the
crystal $|\Delta_{\rm xtl}\phi|$ grows linearly with the width of the
slab. Consequently, the electrostatic contribution to the surface
energy diverges \cite{tasker1979stability,nosker1970polar} as the
width of the crystal increases, and necessitates polarity
compensation. In Fig.~\ref{fig:PolarSlabSchematic}\,(b), we now
consider the crystal immersed in an aqueous environment, e.g. an
electrolyte solution. In this case, Helmholtz layers are established
with surface charge densities $\pm\sigma$, which act to reduce
$|\Delta_{\rm xtl}\phi|$. Under CNC conditions, $\sigma = \sigma_{\rm
  CNC}$, and $|\Delta_{\rm xtl}\phi| = 0$. For large enough crystal
widths, a sufficient number of ions can adsorb to the Helmholtz layer
such that CNC conditions are achieved. For thin crystal widths,
however, incomplete screening occurs, establishing an electric field
across the crystal
\cite{zhang2016finite,sayer2017charge,sayer2019finite} ($|\Delta_{\rm
  xtl}\phi| \neq 0$). If our aim is to model systems on the
macroscopic scale, this poses a severe challenge for molecular
simulations, where one can typically only afford to simulate on the
order of $10^0-10^1$ atomic layers.

The issue of incomplete screening under periodic boundary conditions
(PBC) that are often used in molecular simulations has been the
subject of recent investigations by Zhang \emph{et al}
\cite{zhang2016finite,sayer2017charge,sayer2019finite}. In these
studies, which build upon the work of Stengel, Spaldin and Vanderbilt
\cite{stengel2009electric}, a theoretical framework with which to
model uniform electric and electric displacements fields under PBC has
been established. We refer to these techniques as the \emph{finite
  field methods}. Based on thermodynamic arguments, Zhang and Sprik
showed that the Hamiltonians \cite{zhang2016computing1},
\begin{equation}
  \label{eqn:HD}
  \mathcal{H}_{D}(\mbf{p}^{N},\mbf{r}^{N}) =
  \mathcal{H}_{\rm PBC}(\mbf{p}^{N},\mbf{r}^{N}) + \frac{\Omega}{8\pi}\big(D_z-4\pi P_z(\mbf{r}^N)\big)^2,
\end{equation}
and,
\begin{equation}
  \label{eqn:HE}
  \mathcal{H}_{E}(\mbf{p}^{N},\mbf{r}^{N}) =
  \mathcal{H}_{\rm PBC}(\mbf{p}^{N},\mbf{r}^{N}) - \Omega P(\mbf{r}^N)E_z,
\end{equation}
generate dynamics for a system held under constant electric
displacement field $D_z$ and constant electric field $E_z$,
respectively. Both of these fields are aligned along the surface
normal, which we take to be the $z$ direction in a Cartesian
coordinate system. The momenta and positions of the particles are
denoted by $\mbf{p}^{N}$ and $\mbf{r}^{N}$, respectively, and
$\mathcal{H}_{\rm PBC}$ describes the kinetic energy, and the
potential energy arising from molecular interactions. It is important
to note that the use of 3D Ewald summation with tin-foil boundary
conditions is implicitly assumed in $\mathcal{H}_{\rm PBC}$. The $z$
component of the polarization at time $t$ is given by,
\begin{equation}
  \label{eqn:Pz}
  P_z(t) = \frac{1}{\Omega}\sum_i q_iz_i(t),
\end{equation}
where $\Omega$ is the total volume of the orthorhombic simulation
cell, $z_i$ is the $z$-component of the $i^{\rm th}$ particle's
position, which carries a charge $q_i$. The polarization is defined as
the time integral of the current, and the sum runs over all species in
the system (including free ions). This means that the only source of
electric displacement comes from charges at the `boundaries at
infinity'. It is also important to note that the $z_i$ that enter
Eq.~\ref{eqn:Pz} do not necessarily correspond to the particle's
position in the primary simulation cell; when a particle traverses the
cell boundary, its position is followed out of the cell. This is known
as the \emph{itinerant polarization} \cite{caillol1994comments}. We
also stress that all fields ($D_z$, $E_z$ and $P_z$) that appear in
Eqs.~\ref{eqn:HD} and~\ref{eqn:HE} are uniform, \tcb{and that the
  forces derived from $\mathcal{H}_{E}$ and $\mathcal{H}_{D}$ apply
  both to the solvent/electrolyte, and the \ce{AgI} ions}. The finite
field methods have been used to calculate the dielectric constant of
pure water using both classical \cite{zhang2016computing1} and
\emph{ab initio} molecular dynamics \cite{zhang2016computing2},
\tcr{as well as the conductivities and dielectric constants of aqueous
  electrolyte solutions} \cite{cox2019finite}. They have also been
used to compute the capacitance of the Helmholtz layer at charged
interfaces
\cite{zhang2016finite,sayer2017charge,sayer2019finite,zhang2018communication},
including the polar \ce{NaCl} \hkl(111) surfaces.

Armed with the Hamiltonians given by Eqs.~\ref{eqn:HD}
and~\ref{eqn:HE}, the premise of using the finite field methods to
overcome the necessarily small widths of crystal is simple: If one can
impose a field ($E_z$ or $D_z$) such that $|\Delta_{\rm xtl}\phi| =
0$, then one can force the aqueous environment to provide the
appropriate compensating charge. This was the approach adopted in
Refs.~\onlinecite{zhang2016finite,sayer2017charge,sayer2019finite} to
calculate the capacitance of the Helmholtz layer. In these studies,
the crystal was held fixed. Here we push the argument further and test
whether or not enforcing a compensating charge is sufficient to
stabilize \ce{AgI}'s polar surfaces on timescales relevant to ice
formation. Before pursuing this, however, we first briefly discuss the
finite field methods in comparison to the popular Yeh-Berkowitz (YB)
correction.

The YB correction was developed as a relatively inexpensive procedure
to remove interactions between periodic images along the $z$-direction
in simulations employing a slab geometry
\cite{YehBerkowitz1999sjc}. It works by adding a force $F_{z,i}^{(\rm
  YB)} = -4\pi q_i P_z$ to each particle $i$ in the simulation. It is
straightforward to verify that this is the same force arising from the
second term in Eq.~\ref{eqn:HD} with $D_z=0$. The equivalence of the
$D_z=0$ ensemble and the YB correction has been previously
acknowledged in Ref.~\onlinecite{zhang2016finite}, where it was also
shown that the vacuum spacing normally employed is not a
requirement. In the remainder of this section, we will explain the
procedures for establishing the CNC conditions in the constant $E_z$
and $D_z$ ensembles, and then go on to directly compare results from
simulations performed at $D_z=0$, $D_z = D_{\rm CNC}$ and $E_z =
E_{\rm CNC}$, where $D_{\rm CNC}$ and $E_{\rm CNC}$ are the fields
that impose the appropriate compensating charge. We undertake this
task as the YB correction was explicitly used by Fraux and Doye
\cite{fraux2014note} in their study of ice formation at
\ce{AgI}. Moreover, \tcr{in the supporting information}, we argue that
the `mirrored slab' geometry employed by Zielke \emph{et al}
\cite{zielke2014molecular}, and Glatz and Sarupria
\cite{glatz2016surface} corresponds on average to the $D_z = 0$
ensemble. We will show that use of the $D_z=0$ ensemble has severe
consequences regarding the stability of the crystal. Importantly, in
cases where the slab is held fixed, we find that using $D_z = 0$
rather than $D_{\rm CNC}$ or $E_{\rm CNC}$ has important implications
for the structure of the water at the interface.

\subsection{Establishing the CNC conditions}
\label{subsec:CNC}

Here we briefly overview how $E_{\rm CNC}$ and $D_{\rm CNC}$ are
determined. As the underlying theory has been given in detail
elsewhere \cite{sayer2017charge,sayer2019finite}, we limit ourselves
to highlighting only the most salient aspects relevant to the current
study. \tcr{A more detailed derivation is given in the supporting
  information.} We will work exclusively with the so-called `insulator
centered supercell' geometry (ICS),\footnote{One could also consider
  an `electrolyte centered supercell' (ECS) in which the crystal slab
  straddles the cell boundary. It has been shown previously
  \cite{sayer2019finite} that one can obtain consistent results
  between the ICS and ECS, and so we do not consider the ECS here.}
(\tcr{see Fig.~S3}). In this setup, the length of the simulation cell
is $L_z$, and the primary simulation cell spans $-L_z/2 \le z <
L_z/2$. The \ce{AgI} slab comprises $n+1$ layers of ions, where $n$ is
an odd integer, and is centered around $z=0$. We initially consider a
case where the regions above and below the crystal are filled with an
aqueous electrolyte solution.

\begin{figure}[tb]
  \includegraphics[width=7.65cm]{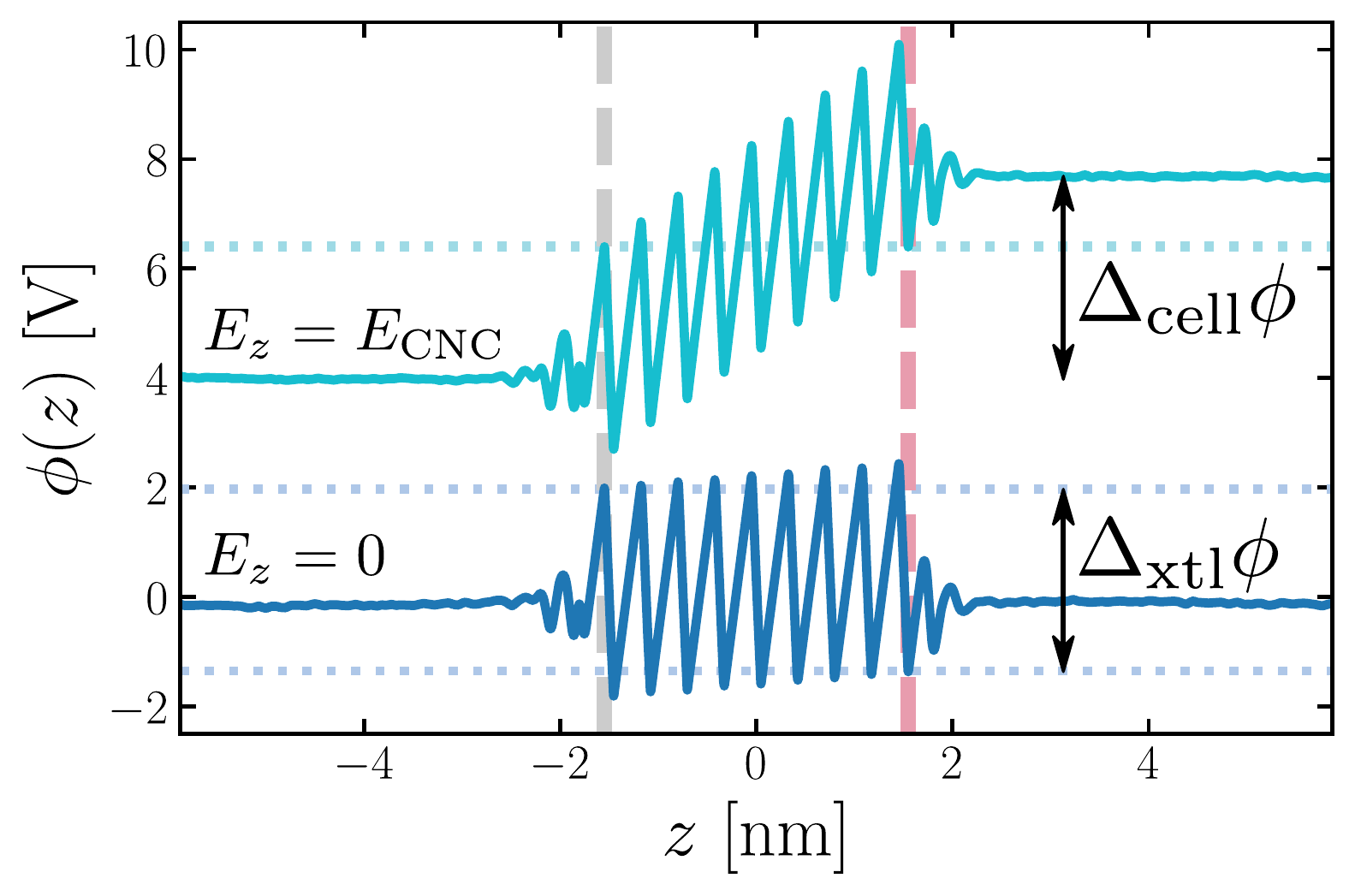}
  \caption{Establishing CNC conditions under constant $E_z$. The solid
    lines show $\phi(z)$ obtained at $E_z = 0$ (dark blue) and $E_{\rm
      CNC} = -0.31$\,V/nm (cyan). For clarity, the latter has been
    shifted up by 6\,V. With $E_z=0$, there is a potential drop
    $\Delta_{\rm xtl}\phi \approx -3.33$\,V across the slab, whereas
    at $E_{\rm CNC}$, there is instead a potential drop across the
    simulation cell, $\Delta_{\rm cell}\phi \approx 3.64$\,V. The
    vertical dashed lines at $\pm 1.55$\,nm indicate the surfaces of
    the crystal. The solution is an aqueous electrolyte, and results
    have been obtained at 298\,K.}
  \label{fig:CNCconditions}
\end{figure}

We begin by considering $E_{\rm CNC}$. The dark blue line in
Fig.~\ref{fig:CNCconditions}\, shows $\phi(z)$ for a \ce{AgI} slab
with $n=17$, obtained from a simulation in which $E_z = 0$. In this
simulation, the crystal was immobile. The location of the
Ag-\hkl(0001) and I-\hkl(000-1) are indicated by dashed lines at
$z_{+} \approx -1.55$\,nm and $z_{-} \approx 1.55$\,nm,
respectively. It is clear there is a potential drop across the slab of
$\Delta_{\rm xtl}\phi \approx -3.33$\,V, corresponding to an average
electric field across the slab of approximately 1.1\,V/nm. $E_{\rm
  CNC}$ can be found empirically by repeating the simulation, but
imposing different values of $E_z$, and measuring $\Delta_{\rm
  xtl}\phi$ in each instance (see \tcr{Fig.~S4}). For this system, we
find $E_{\rm CNC} \approx -0.31$\,V/nm. The resulting $\phi(z)$ is
shown by the cyan line in Fig.~\ref{fig:CNCconditions}. Whereas we
have effectively eliminated $\Delta_{\rm xtl}\phi$, there is now a
potential drop across the simulation cell $\Delta_{\rm cell}\phi
\approx 3.64$\,V. Despite the form of $\phi(z)$, it is important to
realize that the particles do not experience an impulsive force as
they traverse the cell boundary; the field exerts a force $q_iE_z$ on
each particle, irrespective of the particle's position. This can be
seen from Eq.~\ref{eqn:HE}. Note that from the Maxwell relation $D_z =
E_z + 4\pi P_z$, we can obtain an estimate for $D_{\rm CNC}$ by
measuring $\langle P\rangle_{E_{\rm CNC}}$, the average polarization
at $E_{\rm CNC}$. In this instance, we find $\langle D\rangle_{E_{\rm
    CNC}} \approx -14.95$\,V/nm. This can be used as a consistency
check for theoretical predictions of $D_{\rm CNC}$. \tcb{Following the
  symmetry-preserving mean-field theory of Hu \cite{hu2014symmetry},
  Pan \etal\cite{pan2018analytic} have recently derived an analytic
  formula for $E_{\rm CNC}$ for the case of two oppositely charged
  sheets (effectively $n=1$ in the current context), provided one has
  a reasonable estimate of the separation of the Helmholtz layer from
  the crystal. Generalizing such an approach for $n>1$ may prove
  fruitful for future studies.}

We now turn our attention to $D_{\rm CNC}$. While one could take the
approach based on trial-and-error outlined above for $E_{\rm CNC}$,
the $D_z$ ensemble lends itself to a more elegant solution. \tcb{By
  solving a continuum Stern model, it was shown in
  Ref.~\onlinecite{sayer2019finite} that for the ICS, the CNC
  condition is simply,}
\begin{equation}
  \label{eqn:DCNC}
  D_{\rm CNC} = -4\pi\sigma_{\rm CNC},
\end{equation}
where $\sigma_{\rm CNC}$ is the surface charge density of the
Helmholtz layer such that polarity compensation is achieved. By
solving a \tcb{similar} continuum Stern model, we show in the
\tcr{supporting information} that,
\begin{equation}
  \label{eqn:sigmaCNC}
  \sigma_{\rm CNC} = \frac{(n+1)R_1}{(n+1)R_1 + (n-1)R_2}\sigma_0,
\end{equation}
with $\sigma_0$ the surface charge density on each plane of the
crystal, and $R_1$ and $R_2$ are the distances separating the planes
(see Fig~\ref{fig:PolarSlabSchematic}). For the wurtzite structure,
$R_2/R_1 = 3.2$ such that $\lim_{n\to\infty} \sigma_{\rm CNC} \approx
\sigma_0/4$, in agreement with Nosker \etal{} \cite{nosker1970polar}
For the \ce{AgI} crystal with $n=17$ used in our simulations,
Eqs.~\ref{eqn:DCNC} and~\ref{eqn:sigmaCNC} give $D_{\rm CNC} =
-14.99$\,V/nm in good agreement with $\langle D\rangle_{E_{\rm CNC}}
\approx -14.95$\,V/nm obtained above.  Performing a simulation at
$D_{\rm CNC}$, we find $\Delta_{\rm xtl}\phi \approx 0.2$\,V. In the
case of a mobile slab, however, we have found it more robust to find
$D_{\rm CNC}$ empirically from a simulation at $E_{\rm
  CNC}$. \tcr{Table~S1} gives the values of all fields used in our
simulations.

The $E_{\rm CNC}$ and $D_{\rm CNC}$ conditions given above were
derived in the case that the crystal was surrounded by an electrolyte
solution. Given water's ability to screen electric fields almost
entirely, as characterized by its high dielectric constant, it is
natural to ask whether or not pure water is able to provide polarity
compensation. The trial-and-error approach for determining $E_{\rm
  CNC}$ described above provides a means for answering this question
directly. If it is indeed found that water can provide polarity
compensation, will the CNC conditions for the $D_z$ ensemble remain
the same? We argue that the answer is `yes'. In the derivation of the
CNC conditions for the electrolyte \tcr{(see
  Refs.~~\onlinecite{zhang2016finite,sayer2017charge,sayer2019finite}
  and supporting information)}, $D_z$ determines the value of the
surface charge densities at the cell boundaries, and consquently the
surface charge density of the Helmholtz layer. This is a direct
consequence of a uniform polarization in the electrolyte. In the case
of zero ionic strength, there is no longer a Helmholtz layer. Rather,
a single boundary between the solvent and the crystal must provide the
required charge compensation. If we were to observe a uniform solvent
polarization, it stands to reason that as we require the same value of
$\sigma_{\rm CNC}$, then the value of $D_{\rm CNC}$ will be the same
at zero ionic strength as it is for the electrolyte.

\subsection{Comparing $D_z=0$ with $E_{\rm CNC}$ and $D_{\rm CNC}$}
\label{subsec:comparison}

\begin{figure}[tb]
  \includegraphics[width=7.65cm]{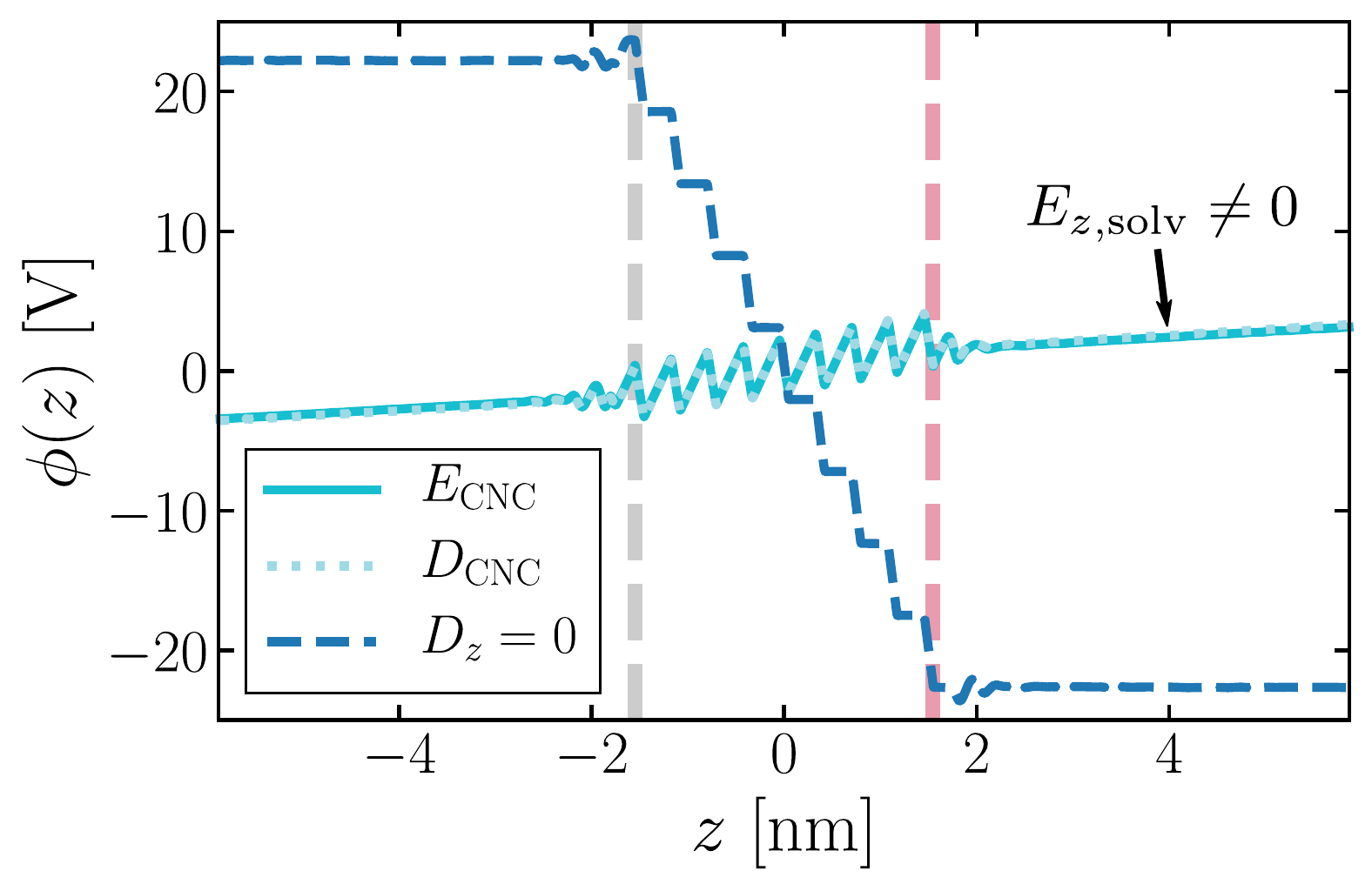}
  \caption{Comparing $\phi(z)$ from different ensembles for \ce{AgI}
    ($n=17$) in contact with pure water at 298\,K. At $D_z=0$ (dashed
    line) there is a large potential drop across the slab,
    $\Delta_{\rm xtl}\phi \approx -46.2$\,V. At $E_{\rm CNC}$ (solid
    line), the potential drop is essentially zero by construction, but
    there is now a finite field in the solvent, $|E_{z,\rm solv}|
    \approx 0.39$\,V/nm. The result obtained at $D_{\rm CNC} =
    -14.99$\,V/nm (dotted line) agrees well with the $E_{\rm CNC}$
    result. Note that the value of $D_{\rm CNC}$ is the same as that
    at finite ionic strength.}
  \label{fig:phi-vs-z_YBcomparison}
\end{figure}

As simulations of heterogeneous ice formation typically consider pure
water in contact with an INP, the prospect of being able to enforce
CNC without ions present is particularly intriguing, as it will permit
a direct comparison of how different electrostatic boundary conditions
affect the crystallization process. To this end, we have found $E_{\rm
  CNC}$ by trail-and-error for an immobile \ce{AgI} crystal in contact
with pure water. In Fig.~\ref{fig:phi-vs-z_YBcomparison}, we show
$\phi(z)$ at $D_z=0$ and $E_z=E_{\rm CNC}$. The result for $D_z=0$ is
striking, with $|\Delta_{\rm xtl}\phi| \approx 46.2$\,V corresponding
to an average electric field of 14.9~V/nm across the slab. On the
other hand, no such large electric field across the crystal is seen at
$E_{\rm CNC}$ (albeit by construction). Rather, what is now observed
is a uniform field in the solvent, $|E_{z,\rm solv}| \approx
0.39$\,V/nm. \tcr{Following our discussion at the end of
  Sec.~\ref{subsec:CNC}}, we therefore expect the value of $D_{\rm
  CNC}$ to still be given by Eqs.~\ref{eqn:DCNC}
and~\ref{eqn:sigmaCNC}. Indeed, we find \tcr{$\langle D\rangle_{E_{\rm
      CNC}} \approx -14.92$\,V/nm} compared to the theoretical
prediction of \tcr{$D_{\rm CNC} = -14.99$\,V/nm}. Performing a
simulation at the theoretical value of $D_{\rm CNC}$ gives $\phi(z)$
shown by the dotted line in Fig.~\ref{fig:phi-vs-z_YBcomparison},
which agrees well with the profile obtained at $E_{\rm CNC}$.

\begin{figure}[b]
  \includegraphics[width=7.65cm]{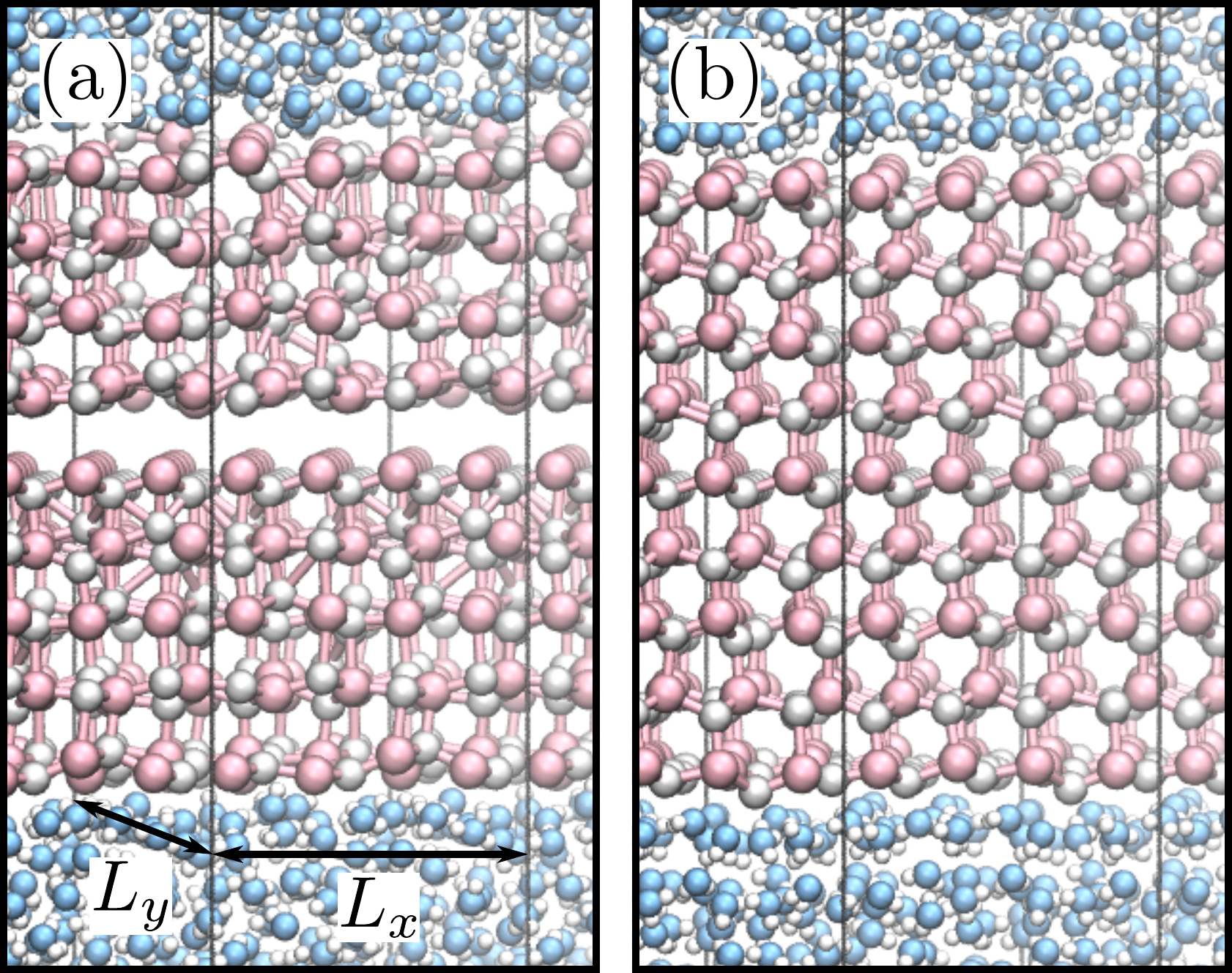}
  \caption{Snapshots from simulations with a mobile \ce{AgI}
    crystal. (a) At $D_z=0$ the \ce{AgI} loses its wurtzite crystal
    structure almost immediately (snapshot taken after 50\,ps). (b) At
    $E_{\rm CNC}$, on the other hand, the \ce{AgI} crystal maintains
    its crystal structure on the nanosecond timescale (snapshot taken
    after 1.6\,ns). In both cases, the central plane of \ce{Ag+} and
    \ce{I-} ions are held fixed. Color scheme: \ce{Ag+}, silver;
    \ce{I-}, pink; \ce{O}, blue; and \ce{H}, white. The black lines
    indicate the simulation cell boundaries. Only part of the
    simulation cell is shown. The solution is pure water, and results
    have been obtained at 298\,K.}
  \label{fig:snaps_composite}
\end{figure}

\begin{figure}[tb]
  \includegraphics[width=7.65cm]{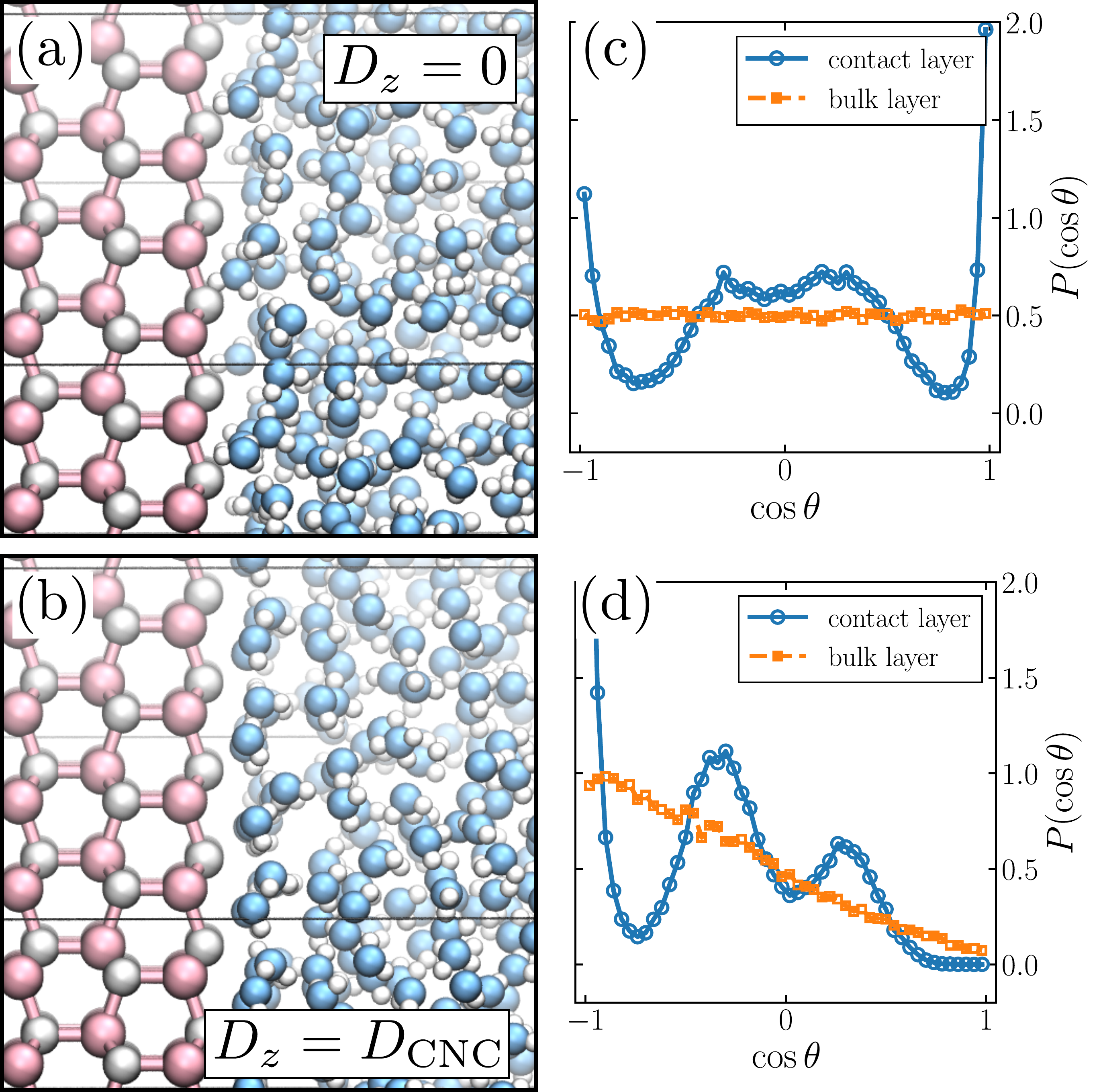}
  \caption{Electrostatic boundary conditions greatly influence the
    structure of water at the interface. (a) and (b) show close up
    snapshots of the \ce{Ag}-\hkl(0001)/\ce{H2O} interface with an
    immobile \ce{AgI} crystal at 298\,K, with $D_{z}=0$ and at $D_{\rm
      CNC}$, respectively. At $D_z = 0$, a significant proportion of
    molecules in the contact layer direct O--H bonds toward the
    positively charged \ce{Ag}-\hkl(0001) surface. In contrast, at
    $D_{\rm CNC}$ no O--H bonds are directed toward the interface, as
    seen in (b). Color scheme as in
    Fig.~\ref{fig:snaps_composite}. Only part of the simulation cell
    is shown. (c) and (d) show $P(\cos\theta)$ at $D_z=0$ and $D_{\rm
      CNC}$, respectively, both for water molecules in the contact
    layer (blue circles) and in bulk solvent (orange squares), where
    differences in structure are also observed. $\cos\theta = +1$ and
    $\cos\theta=-1$ indicate O--H bonds directed immediately toward
    and away from \ce{Ag}-\hkl(0001), respectively.}
  \label{fig:closeups_composite}
\end{figure}

The consequences of such a large field across the crystal with $D_z=0$
are severe. This is demonstrated in
Fig.~\ref{fig:snaps_composite}\,(a), which shows a snapshot from a
$D_z=0$ simulation in which the \ce{Ag+} and \ce{I-} ions are free to
move. After just 50\,ps, the slab no longer resembles the wurtzite
structure of \ce{AgI}. In contrast, at $E_{\rm CNC}$ or $D_{\rm CNC}$,
the crystal remains close to the wurtzite structure, even on the
nanosecond timescale, as shown in Fig.~\ref{fig:snaps_composite}\,(b)
for $E_{\rm CNC}$. While the above results demonstrate the extreme
care required when dealing with polar surfaces like those at \ce{AgI},
it is still common practice to model crystalline lattices in contact
with water as rigid substrates. One may therefore argue that enforcing
CNC conditions by imposing $E_{\rm CNC}$ or $D_{\rm CNC}$ is only of
secondary importance. However, even when using an immobile \ce{AgI}
crystal, the effects on the structure of the water in the contact
layer are profound. This is demonstrated in
Figs.~\ref{fig:closeups_composite}\,(a) and~(b), where we show
snapshots that focus on the contact layer from simulations at $D_z=0$
and $D_{\rm CNC}$, respectively. In the case of the former, we see a
large proportion of water molecules directing O--H bonds toward the
positively charged Ag-\hkl(0001) surface. In contrast, at $D_{\rm
  CNC}$ no O--H bonds are directed toward the interface. These
observations from single snapshots are corroborated by
Figs.~\ref{fig:closeups_composite}\,(c) and~(d), where we show the
probability distribution functions of the O--H bond orientations in
the contact layer obtained from averages over the entire trajectory
(see Sec.~\ref{subsec:BOstats}). Also shown are distributions in the
bulk region in both cases. At $D_z=0$, a uniform distribution of O--H
bond orientations is observed. In contrast, at $D_{\rm CNC}$ there is
a preference for O--H bonds to be directed away from the Ag-\hkl(0001)
surface. This broken symmetry is consistent with $E_{z,\rm solv}\neq
0$ reported in Fig.~\ref{fig:phi-vs-z_YBcomparison}. Below we will
investigate the implications that these differences in interfacial
liquid structure have for ice formation. However, we expect the
behavior observed at the \ce{AgI}/\ce{H2O} interface to be similar at
other polar substrates. Given the widespread use of the YB correction
(or $D_z=0$ ensemble), we expect our findings to be important for the
modeling of a wide variety of other systems too.

\section{Ice formation at silver iodide}

\subsection{Pure water}
\label{subsec:IceNoIons}

By enforcing CNC conditions with the finite field Hamiltonians
(Eqs.~\ref{eqn:HD} and~\ref{eqn:HE}), we have established that the
polar \ce{Ag}-\hkl(0001) and \ce{I}-\hkl(000-1) surfaces are stable in
an aqueous environment, at least on the nanosecond timescale. We have
also observed pronounced differences in the structure of the
interfacial water when simulated at $D_z=0$ and under CNC
conditions. In the absence of ions, however, we also observed a finite
field in the solvent, $|E_{z,\rm solv}| \approx 0.39$\,V/nm. While a
finite electric field inside a dielectric is not a problem in
principle, in practice such a large field would likely lead to the
dielectric breakdown of the water. Nevertheless, as simulations of ice
formation at AgI have typically focused on systems at zero ionic
strength \cite{zielke2014molecular,fraux2014note,glatz2016surface}, it
is instructive to compare and contrast ice formation for pure water in
contact with \ce{AgI} both at $D_z=0$ and at CNC conditions. Moreover,
the pure water system acts as a useful (albeit unphysical) baseline to
help understand the effects of ionic solutes.

\begin{figure}[tb]
  \includegraphics[width=7.65cm]{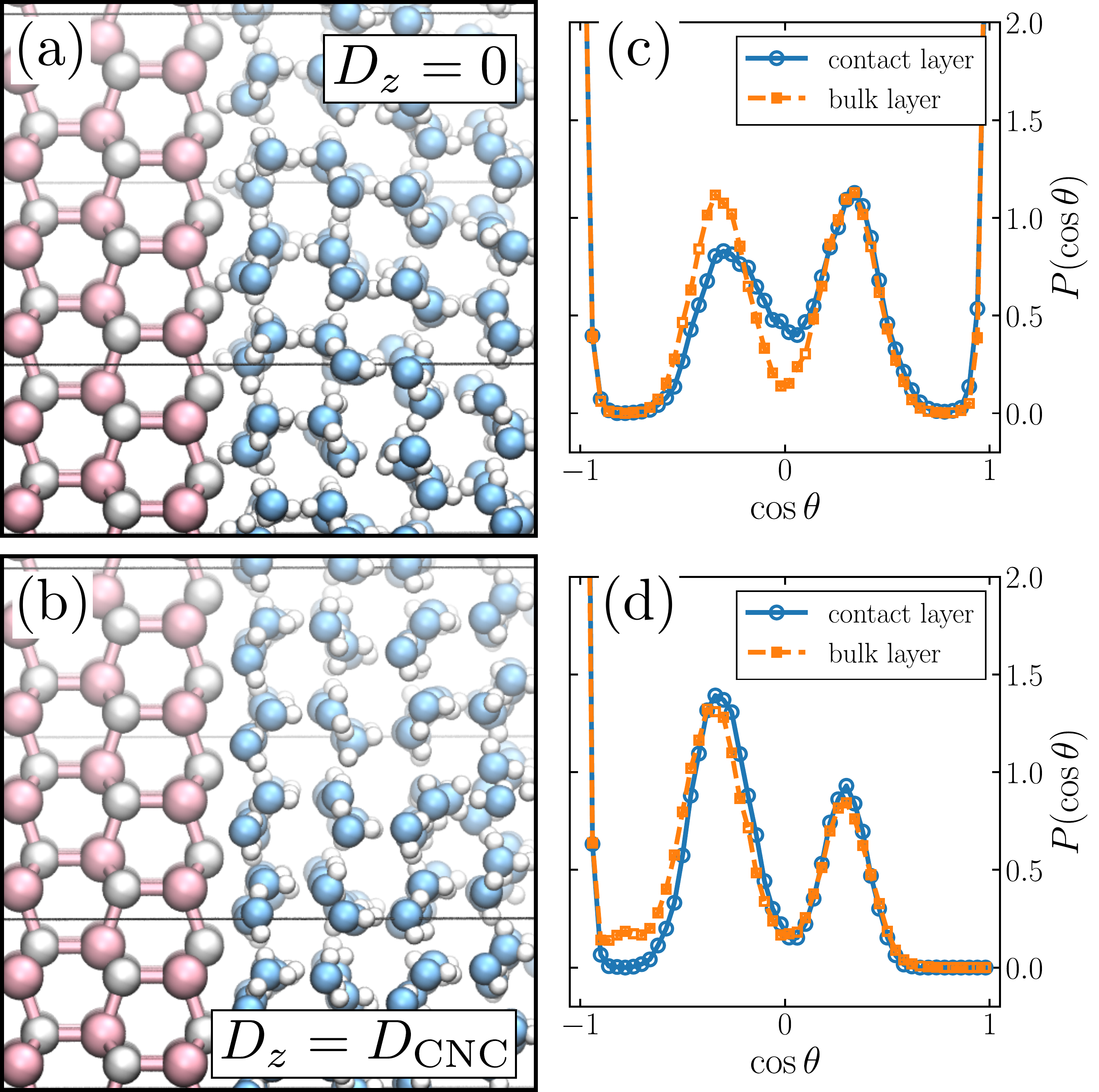}
  \caption{Differences in liquid structure persist upon ice
    formation. Panels (a)-(d) are analogous to those in
    Fig.~\ref{fig:closeups_composite}, but here they are obtained
    after ice formation at 242\,K (results shown for an immobile
    \ce{AgI} crystal). At $D_{z}=0$, O--H bonds are directed both
    toward and away from the interface [(a) and (c)]. In contrast, at
    $D_{\rm CNC}$ no O--H bonds are directed toward Ag-\hkl(0001) [(b)
      and (d)].}
  \label{fig:IceNoIons}
\end{figure}

To investigate ice formation, we adopted the simulation protocol
outlined in Sec~\ref{sec:methods}. For each ensemble (i.e. $D_z=0$,
$D_z=D _{\rm CNC}$ or $E_z=E_{\rm CNC}$), three simulations using this
protocol were performed with an immobile \ce{AgI} crystal. Under CNC
conditions, simulations with a mobile \ce{AgI} crystal were also
performed; as this did not appear to greatly affect the mechanism,
however, \tcr{these results are included in the supporting
  information}. In Fig.~\ref{fig:IceNoIons}\,(a) we show a
representative snapshot of the system after ice formation with
$D_z=0$. Consistent with previous studies, ice is seen to form
preferentially at \ce{Ag}-\hkl(0001) rather than I-\hkl(000-1). This
demonstrates that our simulation setup is sufficiently robust to
capture the general results of previous studies, despite the use of
smaller simulation cells, and a lack of a vacuum gap between periodic
replicas normal to the \ce{AgI} surface. Under CNC conditions, this
preference for ice formation at \ce{Ag}-\hkl(0001) rather than
\ce{I}-\hkl(000-1) persists. However, the occurrence of significant
transient ice-like structures is more pronounced at \ce{I}-\hkl(000-1)
under CNC conditions than it is at $D_z=0$, and indeed, in some of our
simulations ice formation is observed at \ce{I}-\hkl(000-1) as well as
\ce{Ag}-\hkl(0001), \tcr{see Fig.~S11}.

How does the structure of the ice that forms at $D_z=0$ and under CNC
conditions compare? In Figs.~\ref{fig:IceNoIons}\,(a) and~(b) we show
snapshots of the system after ice formation for each ensemble, along
with the corresponding distributions of O--H bond orientations in
Figs.~\ref{fig:IceNoIons}\,(c) and~(d). It is apparent that the
differences in liquid state structure reported in
Fig.~\ref{fig:closeups_composite} greatly influence the structure of
the ice that form. At $D_z=0$ we see O--H bonds directed toward and
away from the interface, both in the contact layer, and in the ice
that forms away from the surface. In contrast, at $D_{\rm CNC}$ no
O--H bonds are directed toward Ag-\hkl(0001).

\subsection{Finite ionic strength}
\label{subsec:IceWithIons}

The `pure \ce{H2O} + \ce{AgI}' system investigated in
Sec.~\ref{subsec:IceNoIons} provides an interesting comparison study
of the $D_z=0$ and CNC ensembles. Nevertheless, in both instances
there are unphysical aspects. At $D_{z}=0$ it is not possible to
simulate the crystal with mobile \ce{Ag+} and \ce{I-} ions owing to a
large potential drop across the slab. Conversely, under CNC conditions
there is an unrealistically large electric field in the solvent. This
is strong motivation to investigate the effects of ions on ice
formation, as such mobile charge carriers may provide polarity
compensation while maintaining zero electric field far from the
crystal (see Fig.~\ref{fig:CNCconditions}). Here we will restrict
ourselves to a simple \ce{NaCl} aqueous electrolyte for which
reasonable simple point charge models are readily available
\cite{benavides2017potential}. However, we emphasize that using the
finite field methods to enforce CNC conditions can be readily applied
to other systems too. As it is known experimentally that ions affect
ice formation in nontrivial ways---both at \ce{AgI}
\cite{reischel1975freezing} and other
surfaces\cite{whale2018enhancement,kumar2018ice}---the work presented in this
section serves as a platform from which to study ice formation in more
complex electrolytes.

\begin{figure}[tb]
  \includegraphics[width=7.65cm]{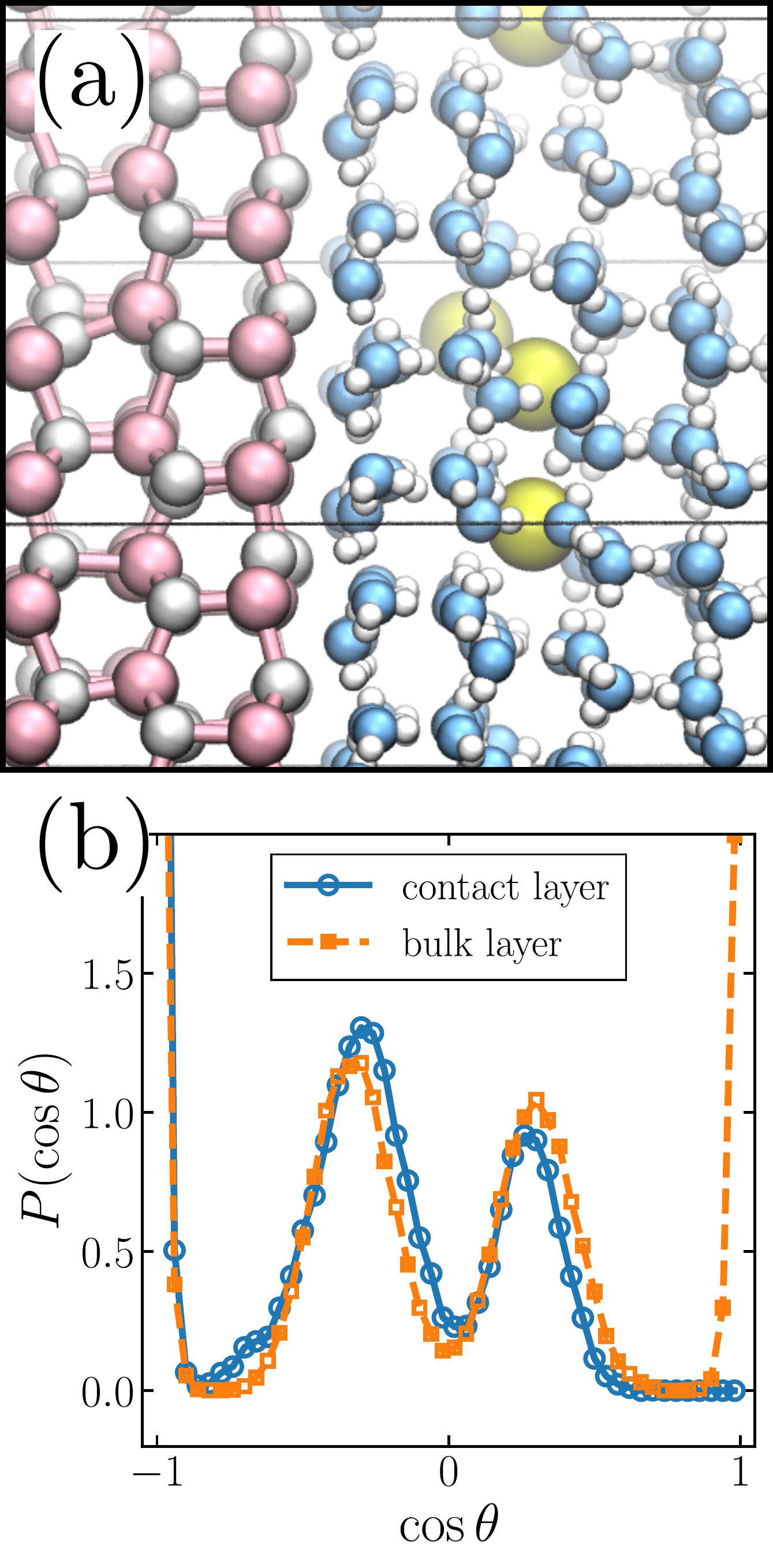}
  \caption{Ice formation in the presence of ions gives rise to a
    proton ordered contact layer, but proton disorder away from the
    surface. (a) Close up snapshot after ice formation at
    Ag-\hkl(0001) at 242\,K, with a mobile \ce{AgI} slab at $D_{\rm
      CNC}$. (b) $P(\cos\theta)$ both for the contact layer and a
    layer in the bulk. The \ce{Cl-} ions are shown in yellow,
    otherwise the color scheme is the same as
    Fig.~\ref{fig:snaps_composite}.}
  \label{fig:ions_closeup}
\end{figure}

For the $E_{\rm CNC}$ and $D_{\rm CNC}$ ensembles, we simulated ice
formation using the same protocol as for the pure water system (see
Sec.~\ref{sec:methods}). In order to mitigate colligative effects, we
decided to simulate three ion pairs, which is in principle sufficient
to enforce CNC conditions (Eq.~\ref{eqn:sigmaCNC}). In
Fig.~\ref{fig:ions_closeup}\,(a) we show a snapshot after ice
formation has occurred at $D_{\rm CNC}$ in the presence of a mobile
\ce{AgI} slab. As in the case without ions, ice formation is still
observed to occur preferentially at \ce{Ag}-\hkl(0001) rather than
\ce{I}-\hkl(000-1). However, while the structure of the water in the
contact layer is similar to that seen in the absence of ions, it is
now clear that this structure is lost further from the interface. This
is shown quantitatively by the probability distribution functions of
O--H bond orientations in Fig.~\ref{fig:ions_closeup}\,(b). By acting
as hydrogen bond acceptors, it appears that the \ce{Cl-} ions
sufficiently disrupt the polar hydrogen bond network found under CNC
conditions in the pure water case.

Finally, it is natural to ask about the effects of ions on the
kinetics of ice formation. Given the small simulation cells and the
limited number of simulations performed (three for each set of
conditions), we are not in a position to make firm statements in this
regard. Nevertheless, it does appear that ice formation is generally
slower in the presence of dissolved ions, and undergoes a mechanism
more akin to traditional nucleation i.e. a long induction time
followed by relatively rapid crystal growth \tcr{(see Figs.~S9
  and~S10)}. These differences are particularly pronounced when
compared to the $D_z=0$ ensemble results, where crystal formation
appears especially fast. We also performed a set of simulations at
$E_{\rm CNC}$ with a mobile \ce{AgI} slab but with the signs of the
dissolved ions swapped i.e. a hypothetical ``\ce{Na-} + \ce{Cl+}''
system. In this case, no ice formation was observed on the time scale
of our simulations \tcr{(approx. 350\,ns)}. This null result indicates
that ion specific details are indeed important for ice formation, and
that the role of the ions extends beyond simply providing mobile
charge to stabilize the surface.

\section{Concluding Remarks}
\label{sec:concl}

The focus of this article has been whether or not the polar
\ce{Ag}-\hkl(0001) and \ce{I}-\hkl(000-1) surfaces of \ce{AgI} are
stable in aqueous solution on timescales relevant to ice formation. To
achieve this, we have exploited recent advances in simulation
methodology \cite{zhang2016computing1,zhang2016finite,sprik2018finite}
that enable us to enforce conditions of compensating net charge, thus
ensuring that the drop in electrostatic potential across the crystal
vanishes. This is a necessary condition for a finite surface free
energy. We have found that under CNC conditions, the polar surfaces of
\ce{AgI} are indeed sufficiently stable to facilitate ice
formation. Importantly, however, we have also found that the presence
of dissolved ions is crucial in this regard; without these mobile
charge carriers there exists a finite electric field in the aqueous
phase. For the systems studied here, the magnitude of this field is
unrealistically large. More generally, a finite uniform electric field
will engender stability issues as the thickness of the liquid layer
increases, in a similar manner to thin film polar oxides
\cite{noguera2000polar,goniakowski2008polarity}. For macroscopic
samples sizes, we conclude that the presence of mobile charge carriers
is paramount for stability.

As discussed in the introduction, we have only considered a polarity
compensation mechanism by which the aqueous environment supplies the
required compensating charge, and we have neglected the possibility of
electronic and nonstoichiometric reconstruction. This was motivated in
part by the long held view that the close structural similarity
between \ce{AgI} and ice is the cause of its excellent ice nucleating
properties \cite{pk97}. The results presented here indeed suggest that
this is a plausible explanation, although complicated by the polar
surfaces' need for proximate dissolved ions. While we cannot preclude
electronic and nonstoichiometric reconstruction, a thorough study of
the latter would likely require the development of improved force
fields, while the former would call for explicit calculation of the
electronic structure. These lie beyond the scope of the current
article. Ultimately, the relative importance of these different
mechanisms will be determined by the relative free energies and
kinetic barriers separating the appropriate states. Enforcing CNC
conditions in the presence of the aqueous environment will at the very
least provide an appropriate reference state.

For pure water in contact with \ce{Ag}-\hkl(0001) and
\ce{I}-\hkl(000-1) we also compared to simulations performed at
$D_z=0$, which has the same Hamiltonian as the commonly used
Yeh-Berkowitz method \cite{YehBerkowitz1999sjc}. We found the contrast
with the system under CNC conditions to be stark: At $D_z=0$ a
significant proportion of O--H bonds were found to be directed toward
the positively charged \ce{Ag}-(0001) surface, whereas under CNC
conditions no O--H bonds were found to point at the surface. This
difference in contact layer structure was seen to persist upon
introduction of dissolved ions. We expect this result to have
implications beyond the \ce{AgI} system considered here. It is worth
emphasizing that to enforce CNC conditions we have used two different
methods: Imposing a uniform electric field, or imposing a uniform
electric displacement field. The equations of motion for these two
ensembles are different, and correspond to distinctly different
electrostatic boundary conditions
\cite{zhang2016computing1,sprik2018finite}. It is
therefore rather satisfying that results obtained at $D_{\rm CNC}$ and
$E_{\rm CNC}$ are broadly in agreement with each other.


Let us put this work in the context of ice nucleation more
broadly. Throughout this study we have used relatively small
simulation cells and ``off-the-shelf'' non-polarizable force
fields. These have been sufficient for the purpose of demonstrating
the effects of different electrostatic boundary conditions on the
stability of \ce{Ag}-\hkl(0001) and \ce{I}-\hkl(000-1) in aqueous
environments, and the potential impact this has for ice formation. To
obtain quantitative kinetic data would require the use of much larger
simulations in combination with e.g. seeding techniques
\cite{sanz2013homogeneous,espinosa2014homogeneous,pedevilla2018heterogeneous}
or forward flux sampling
\cite{haji2015direct,bi2016heterogeneous,cabriolu2015ice,bi2017enhanced,sosso2016ice,sosso2018unravelling}
to compute rates, which should be readily compatible with the
Hamiltonians given by Eqs.~\ref{eqn:HD} and~\ref{eqn:HE}. The finite
field methods used here can therefore be viewed as an additional tool
for those investigating heterogeneous ice nucleation with computer
simulation. Given it is becoming increasingly apparent that ions
impact heterogeneous ice nucleation in complex ways
\cite{whale2018enhancement,kumar2018ice}, these techniques are likely to be
important for many future studies in this area. Perhaps most
importantly, what our results highlight is the crucial role ions can
play in the heterogeneous ice formation mechanism itself, and should
not be considered as a small perturbation to the water/solid
interface.

\section{Methods}
\label{sec:methods}

\subsection{Force fields and molecular models}
\label{subsec:models}

To model \ce{AgI} we used a reparametrized version
\cite{shimojo1991molecular,bitrian2008molecular} of the
Parrinello-Rahman-Vashista (PRV) force field
\cite{parrinello1983structural}. Non-electrostatic interactions were
computed from a table, which gives consistent results with
Ref.~\onlinecite{bitrian2008molecular} for molten \ce{AgI} \tcr{(see
  Fig.~S18)}. To model water we used the \textsmaller{TIP4P/2005}
model \cite{abascal2005general}, which has a melting temperature
$T_{\rm m} = 252$\,K. For sodium chloride we used the recently
developed Madrid model \cite{benavides2017potential}, whose
non-electrostatic interactions with water are of a simple
Lennard-Jones (LJ) form. This model was designed specifically for use
with \textsmaller{TIP4P/2005}, and gives a good description of the
solubility of \ce{NaCl} in water. With appropriate signs, silver and
iodide ions carried a charge $q_{\ce{AgI}} = 0.5815\,e$, while sodium
and chloride ions carried a charge $q_{\ce{NaCl}} = 0.85\,e$. Despite
the use of these partial charges, for ease of notation we still refer
to these ions as `\ce{Ag+}' etc. throughout the article. Oxygen atoms
of the water molecule carried a charge $q_{\rm O} = -1.1128\,e$ and
the charge on the hydrogen atoms was $q_{\rm H} = -q_{\rm
  O}/2$. Following Fraux and Doye \cite{fraux2014note}, who also used
the \textsmaller{TIP4P/2005} model in their study of ice formation at
\ce{AgI}, the non-electrostatic interactions between the \ce{AgI} ions
and the water molecules were described by a LJ potential centered on
the oxygen atoms of the water molecules, using parameters originally
from Hale and Keifer \cite{hale1980studies}. Lorentz-Berthelot mixing
rules were applied to obtain non-electrostatic interactions between
\ce{NaCl} and \ce{AgI}. \tcr{Parameters for non-electrostatic
  interactions are reported in Tables~S2 and~S3.}

Following Zielke \emph{et al} \cite{zielke2014molecular}, we used
Burley's lattice parameters ($a=0.4592$\,nm, $c=0.7510$\,nm) for
\ce{AgI} \cite{burley1963structure}. All simulations used in this work
comprised $n+1=18$ layers of \ce{AgI}, with each layer itself
comprising 16 \ce{Ag+} or \ce{I-} ions. With the crystal held fixed,
this resulted in a slab width of 3.0934\,nm. The lateral dimensions of
the simulation cell were $L_x=1.8368$\,nm and $L_y=1.5907$\,nm,
resulting in a formal charge density on each layer of $\sigma_0
\approx 3.18\,e/\text{nm}^2$. The total length of the simulation cell
in the $z$-direction (which we take to be normal to the surface) was
$L_z = 11.7475$\,nm. The remaining volume not occupied by \ce{AgI}
contained 750 water molecules, resulting in a number density in the
bulk fluid region of $\rho_{\rm w} \approx 30.4\,\text{nm}^{-3}$ at
298\,K. This is slightly lower than the density of bulk liquid water,
and has been chosen as the finite field methods have been formulated
strictly in the canonical ensemble
\cite{zhang2016computing1,sprik2018finite}; using this lower density
therefore allows enough space for the growing ice crystal. This is
similar to the approach adopted by Zielke \emph{et al}
\cite{zielke2014molecular}. We note that, in contrast, Fraux and Doye
used liquid films with one side in contact with \ce{AgI} and the other
in contact with vacuum, effectively holding the fluid at zero
pressure. As our results without dissolved ions at $D_z=0$ appear
broadly consistent with Fraux and Doye, it suggests the general
features of ice formation at \ce{AgI} are fairly robust to such
simulation details. For simulations with dissolved ions, three
\ce{NaCl} ion pairs were placed in the fluid region, with no further
adjustments to the simulation set up.

\subsection{Simulation protocols}
\label{subsec:protocols}

We have performed two types of simulations for the system described
above. First, we have performed simulations at 298\,K (i.e. water in
the liquid state) in order to establish the CNC conditions (see
Sec.~\ref{subsec:CNC}). Then, we have performed simulations with a
protocol described below to observe ice formation. Throughout this
article we used the \textsmaller{LAMMPS} simulation package
\cite{plimpton1995sjc}, suitably modified to propagate dynamics in the
constant $E_z$ and $D_z$ ensembles with the \textsmaller{TIP4P/2005}
water model. The velocity Verlet algorithm was used to propagate
dynamics with a time step of 2\,fs. To maintain the rigid geometry of
the \textsmaller{TIP4P/2005} water molecules, we used the
\textsmaller{RATTLE} algorithm \cite{andersen1983rattle}. Temperature
was maintained using a Nose-Hoover thermostat
\cite{shinoda2004rapid,tuckerman2006liouville} with damping constant
0.2\,ps. The particle-particle particle-mesh Ewald method was used to
account for long-ranged interactions \cite{HockneyEastwood1988sjc},
with parameters chosen such that the root mean square error in the
forces were a factor $10^{5}$ smaller than the force between two unit
charges separated by a distance of 1.0\,\AA \cite{kolafa1992cutoff}.

For simulations performed at 298\,K, at least 100\,ps of equilibration
was performed, followed by a further 1.5\,ns of production. To compute
the electrostatic potential profiles $\phi(z)$, the procedure outlined
in Ref.~\onlinecite{wirnsberger2016non} was followed. To investigate
ice formation, we used the following protocol. First, dynamics were
propagated at 252\,K for 5\,ns. Then the system was cooled at a rate
of 0.5\,K/ns for 20\,ns to a target temperature of 242\,K. Finally,
the dynamics of the system were propagated at 252\,K until ice
formation was observed, or \tcr{470}\,ns had occurred (whichever was
sooner). Aside from the simulations in which we reversed the signs of
the dissolved ions' charge (see Sec.~\ref{subsec:IceWithIons}),
\tcr{ice formation was observed in all but one simulation}.

\subsection{Bond orientation statistics}
\label{subsec:BOstats}

To quantify the bond orientation statistics at the interface, we have
calculated $\cos\theta$, where $\theta$ is the angle formed between
the O--H bond and the $z$-axis of the simulation cell. Specifically,
if we denote the unit vector pointing from the oxygen atom of a water
molecule to one of its hydrogen atoms (the procedure is repeated for
the other hydrogen) as $\hat{\mbf{b}}$ and the unit vector along the
$z$-direction as $\hat{\mbf{z}}$, then what we in fact calculate is
$\hat{\mbf{b}}\cdot\hat{\mbf{z}} = \cos\theta$. In our simulation
setup, the surface normal of \ce{Ag}-(0001) points along
$-\hat{\mbf{z}}$, thus $\cos\theta=-1$ corresponds to an O--H bond
directed away from the surface, and $\cos\theta=+1$ means an O--H bond
is directed toward the surface. At I-\hkl(000-1) the situation is
reversed, that is, $\cos\theta=-1$ corresponds to an O--H bond
directed toward the surface, and $\cos\theta=+1$ means an O--H bond is
directed away from the surface (\tcr{see Fig.~S12}).

\begin{acknowledgments}
  Thomas Whale and Michiel Sprik are thanked for many helpful
  discussions. Chao Zhang is thanked for reading a draft of the
  manuscript. T.S. is supported by a departmental studentship
  (No. RG84040) sponsored by the Engineering and Sciences Research
  Council (EPSRC) of the United Kingdom. We are grateful for
  computational support from the UK Materials and Molecular Modelling
  Hub, which is partially funded by EPSRC (EP/P020194), for which
  access was obtained via the UKCP consortium and funded by EPSRC
  grant ref EP/P022561/1. S.J.C. is supported by a Royal Commission
  for the Exhibition of 1851 Research Fellowship.
\end{acknowledgments}

\bibliography{../../cox}

\end{document}